# DESIGNING THEORY-DRIVEN ANALYTICS-ENHANCED SELF-REGULATED LEARNING APPLICATIONS

Mohamed Amine Chatti, University of Duisburg-Essen, Germany, mohamed.chatti@uni-due.de

Volkan Yücepur, University of Duisburg-Essen, Germany, volkan.yuecepur@stud.uni-due.de

Arham Muslim, National University of Sciences and Technology, Pakistan, arham.muslim@seecs.edu.pk

Mouadh Guesmi, University of Duisburg-Essen, Germany, mouadh.guesmi@stud.uni-due.de

Shoeb Joarder, University of Duisburg-Essen, Germany, shoeb.joarder@stud.uni-due.de

Abstract:      There is an increased interest in the application of learning analytics (LA) to promote self-regulated learning (SRL). A variety of LA dashboards and indicators were proposed to support different crucial SRL processes, such as planning, awareness, self-reflection, self-monitoring, and feedback. However, the design of these dashboards and indicators is often without reference to theories in learning science, human-computer interaction (HCI), and information visualization (InfoVis). Moreover, there is a lack of theoretically sound frameworks to guide the systematic design and development of LA dashboards and indicators to scaffold SRL. This chapter seeks to explore theoretical underpinnings of the design of LA-enhanced SRL applications, drawing from the fields of learning science, HCI, and InfoVis. We first present the Student-Centered Learning Analytics-enhanced Self-Regulated Learning (SCLA-SRL) methodology for building theory-driven LA-enhanced SRL applications for and with learners. We then put this methodology into practice by designing and developing LA indicators to support novice programmers' SRL in a higher education context.

## 1.      INTRODUCTION

Self-regulated learning (SRL) is one of the most important areas of research within educational psychology over the last two decades. SRL refers to self-generated thoughts, feelings, and behaviors that are oriented to attaining learning goals (Zimmerman, 2000). Self-regulation researchers attributed individual differences in learning to students' lack of self-regulation and provided methods to help students develop key SRL processes, such as goal setting, adoption of learning strategies to attain the goals, self-evaluation, time management, and seeking help or information (Zimmerman, 2002). In recent years, there has been a growing interest regarding the role of learning analytics (LA) to support how students regulate their own learning processes (Winne, 2017). LA aims at turning educational data into insights, decisions, and actions to improve learning. In LA systems, data is traditionally displayed through dashboards with indicator visualizations developed from traces that learners leave when they interact with different learning spaces. Various LA dashboards and indicators were proposed to support crucial SRL processes, such as planning, awareness, self-reflection, self-monitoring, feedback, and motivation, (Bodily & Verbert, 2017; Jivet et al., 2018; Schwendimann et al., 2016). However, current design of LA dashboards and indicators to support SRL suffer from two main limitations. First, the design of the dashboards and indicators is often without reference to SRL theories and models (Gašević et al., 2019;

Jivet et al., 2018). Second, the designed dashboards and indicators are often not well aligned with learners' needs and expectations. In fact, there is still a divide between those who design LA dashboards (i.e. researchers, developers), and those who are expected to use these dashboards or are most affected by them (i.e. learners) (Chatti et al., 2020a; Ifenthaler & Schumacher, 2016). In order to design LA dashboards and indicators that meet learner's needs and expectations, it is crucial that LA researchers and developers conduct qualitative user studies (e.g. interviews, focus groups) where they go to the learners, observe their activities, and try to understand what they really need to support their SRL activities (Jivet et al., 2020; Schumacher & Ifenthaler, 2018). Thereby, it is important to find the right set of questions to ask in the study in order to generate the right set of indicators. To get at this, there is a need to draw on SRL theories and models as a theoretical foundation to guide the design of the qualitative user studies. The next step would be to design the indicators themselves. Here also, having the learners in the loop and empowering them to take control of the indicator design process is crucial in order to effectively meet their needs and expectations.

In the LA research community, there is a lack of theoretically sound methodologies to guide the systematic design and development of LA indicators to scaffold SRL. Several methodologies are established in different disciplines to support the systematic design of user interfaces (Norman, 2013), information visualizations (Munzner, 2014), visual analytics interfaces (Thomas & Cook, 2006), and information systems (Peffers et al., 2007). However, these methodologies are not enough in the LA discipline because they do not take account of the learning context (Martinez-Maldonado et al., 2015). While LA researchers have long been interested in learning theory-driven design of LA dashboards, most of this work has been conducted in conceptual terms (Jivet et al., 2018). And, despite the fact that research on human-centered learning analytics (HCLA) has been gaining momentum in recent years, approaches that involve learners in the design of LA indicators remain rare in the literature (Ochoa & Wise, 2020). Thus, there is a need for a complete methodology that provides us with some guidance about how to conduct theory-driven LA-enhanced SRL research.

In this paper, we argue that in order to design LA dashboards and indicators that effectively support SRL activities, it is essential to (1) **understand** learners' needs and expectations from LA-enhanced SRL and (2) **empower** learners to steer the indicator design process. The guiding questions for this work are: Which LA indicators are needed to support SRL? and How to systematically design these indicators? To answer these questions, we propose and develop a **S**tudent-**C**entered **L**earning **A**nalytics-enhanced **S**elf-**R**egulated **L**earning (**SCLA-SRL**) methodology that provides a process model to conduct theory-driven research on LA-enhanced SRL. The primary aim of SCLA-SRL is to guide the systematic design of LA indicators that support SRL by (1) linking the design to well-established SRL models as well as human-computer interaction (HCI) and information visualization (InfoVis) guidelines and (2) actively involving learners throughout the entire indicator design process. The SCLA-SRL methodology is illustrated with a case study of the systematic design and development of LA indicators to support novice programmers' SRL in a higher education context.

The paper proceeds as follows: The next section provides an overview of the current related work on LA-enhanced SRL, LA dashboards, and student-centered LA. Then, we present and discuss our methodology for Student-Centered Learning Analytics-enhanced Self-Regulated Learning (SCLA-SRL)**.** Next, we demonstrate an example of SCLA-SRL application by utilizing it in a higher education SRL scenario. Finally, we conclude by summarizing the contributions of the work.

## 2.        RELATED WORK

This work aims at providing a methodology to guide the systematic design and development of LA indicators that support SRL. So far, no complete, generalizable process model exists for LA-enhanced SRL research; however, if we develop such a process model, it should build upon the strengths of prior efforts. There is a substantial body of research, both within the LA literature and in related disciplines, that provides us with principles, practices, and procedures to support such a process. Below, we present this related work and discuss how our proposed methodology for conducting LA-enhanced SRL research builds on this work and integrates its principles, practices, and procedures.

## 2.1    SRL Meets LA

Self-regulated learning (SRL) is generally defined as "an active, constructive process whereby learners set goals for their learning and then attempt to monitor, regulate, and control their cognition, motivation, and behavior, guided and constrained by their goals and the contextual features in the environment" (Pintrich, 2000). It includes the cognitive, metacognitive, behavioral, motivational, and emotional/affective aspects of learning (Panadero, 2017). Panadero (2017) provides an excellent analysis and comparison of different SRL models (e.g., Boekaerts, 1992; Efklides, 2011; Pintrich, 2000; Winne & Hadwin, 1998; Zimmerman, 2000). The author points out that although these models address different aspects and use different terminologies, all of them view SRL as a cyclical process, composed of three main phases: (a) *goal setting* (forethought, task analysis, planning, activation of goals, self-motivation); (b) *executing* (performance, processing); and (c) *evaluating* (self-reflection, feedback, monitoring, controlling, appraisal, regulating, adapting, reacting).

Recently, there is an increased interest in the application of LA to promote SRL. Several researchers stressed the need to bring LA into SRL research. For instance, Roll and Winne (2015) pointed out that LA offers exciting opportunities for analyzing and supporting SRL. According to the authors, LA can provide affordances and interventions for learners to more productively regulate their learning. Winne and Baker (2013) also stressed that educational data mining (EDM) - a research field closely related to LA - can play a significant role to advance research on motivation, metacognition, and SRL. Bodily et al. (2018) linked student-facing LA dashboards (LADs) to open learner models (OLMs), as both have similar goals. The authors further stated that OLMs can be used as awareness tools to help learners monitor, reflect on, and regulate their own learning.

Winne (2017) noted that a framework is useful to conceptualize LA for SRL. There are few examples of case studies that integrate LA and SRL by following a theory-driven approach. For example, Nussbaumer et al. (2015) pointed out that LA can provide personalized scaffolds that assist learners in a self-regulated manner. Building on SRL theory, the authors designed and implemented an architecture composed of different learning methodologies for supporting students' SRL in a variety of activities. Similarly, Marzouk et al. (2016) adopted self-determination theory (SDT) as a framework for designing LA that promote SRL as a function of content studied, reasons to adapt learning processes, and the presentation of analytics. Molenaar et al. (2020) presented a learning path app that combines three personalized visualizations to support young learners' SRL in adaptive learning technologies, following the COPES model as a theoretical basis (Winne & Hadwin, 1998).

Chatti and Muslim (2019) pointed out that, while SRL theory has been used to inform the design of LA tools, there remain important gaps in the theory from which to conduct research on LA-enhanced SRL in a systematic manner. Particularly, there is a lack of theoretically sound frameworks to guide the systematic design and development of LA indicators to promote SRL. To address this challenge, the authors proposed the personalization and learning analytics (PERLA) framework that provides a process model to guide the design of qualitative user studies attempting to collect requirements for LA indicators that can support different SRL processes. The proposed framework, however, is at the conceptual level and is still not applied and validated in a real learning setting. In this paper, we build on the PERLA framework and augment it with another process model enabling to move from requirement elicitation to the concrete design and development of LA indicators that support SRL.

## 2.2        Learning Analytics Dashboards

A variety of dashboards presenting data to various LA stakeholder groups were proposed in the LA literature (Jivet et al., 2018; Verbert et al., 2013). LA dashboards (LADs) are "single displays that aggregate different indicators about learners, learning processes and/or learning contexts into one or multiple visualisations" (Schwendimann et al., 2016). They aim at supporting students and teachers in making informed decisions about the learning and teaching process (Jivet et al., 2020). Current reviews of LAD research (e.g., Bodily & Verbert, 2017; Schwendimann et al., 2016; Verbert et al., 2014) tried to identify what data is presented to different LA stakeholders, why is the data presented, and how data can be presented to support sense-making processes.

Research around student facing LADs traditionally has a strong focus on visualizing data to support different crucial SRL processes, such as goal setting and planning (Jivet et al., 2020), (self-)monitoring (Molenaar et al., 2020; Schwendimann et al., 2016), awareness and reflection (Ahn et al., 2019; Bodily & Verbert, 2017; Jivet et al., 2017), metacognition (Bodily et al., 2018; Jivet et al, 2018; Karaoglan Yilmaz & Yilmaz, 2020), and feedback (Molenaar et al., 2020). However, little attention has been paid to the systematic design of the LA indicators to support the intended goals. Indicators represent a core part of any dashboard. An indicator can be defined as "a specific calculator with corresponding visualizations, tied to a specific question" (Muslim et al., 2017). In general, the current design of LA indicators suffers from two main limitations; thus hindering their acceptance and adoption. First, the design of the indicators is rarely grounded in learning theories (Chatti & Muslim, 2019; Gašević et al, 2015; 2019; Jivet et al., 2018; Kelly et al., 2015; Marzouk et al., 2016; Molenaar et al., 2020; Sedrakyan et al., 2018), human-computer interaction (HCI) (Verbert et al., 2013), and information visualization (InfoVis) (Alhadad 2018; Chatti et al., 2020b; Ritsos & Roberts, 2014). Second, the indicators are designed about and not with their users and are thus often not well aligned with user needs. Consequently, users do not see the added-value of these indicators (Chatti et al., 2020a; de Quincey et al., 2019). For example, Jivet et al. (2020) noted that in many cases what students report as relevant to see and use on a dashboard differs from what LADs provide. Thus, there is a crucial need for a framework that (1) draws from existing theories, principles, and practices to inform the design of useful LA indicators that support SRL and (2) put learners in the driver's seat and actively involve them in the design of the indicators that really meet their needs.

## 2.3        Student-Centered Learning Analytics

LA research has recently begun to investigate how HCI principles can be adopted and adapted to support the development of human-centered learning analytics (HCLA) as the key to user acceptance and adoption of LA systems (Chatti et al., 2020a; Buckingham Shum et al., 2019). HCLA is an approach that emphasizes the human factors in LA and aims at bringing HCI to LA in order to involve users throughout the whole LA process (Chatti et al., 2020a). In recent years, some researchers succeeded in bringing design thinking and human-centered design (HCD) into the LA research community. Design thinking is an HCI approach to problem forming and solving that is focused on who we are designing for. HCD - a powerful tool for design thinking - is a user-centered process that starts with the user, creates design artifacts that address real user needs, and then tests those artifacts with real users (Norman, 2013). However, research on HCLA is still in the early stages of development (Buckingham Shum et al., 2019).

In the LA literature to date, there are only few papers, which provide mature examples of how HCI approaches (e.g., design thinking, HCD, participatory design, value-sensitive design) can be applied to LA. These works mainly present case studies that target teachers or institutions as end-users (e.g., Ahn et al., 2019; Dollinger et al, 2019; Holstein et al., 2019; Martinez-Maldonado et al., 2015; Rehrey et al., 2019). Examples of HCLA research involving learners within the design process remain scarce (e.g., de Quincey et al., 2019; Prieto-Alvarez et al., 2018; 2020; Sarmiento et al, 2020). In addition, the reported case studies are focused on the participatory design of LA tools and platforms (macro design level) rather than the systematic design of the underlying indicators (micro design level). Furthermore, most of the current work on HCLA is not informed by well-established InfoVis design guidelines. To fill this gap, Chatti et al. (2020a) proposed the Human-Centered Indicator Design (HCID) as an HCLA approach that targets learners as end-users by involving them in the systematic design of LA indicators that fit their needs, based on conceptual models from the HCI and InfoVis fields. This approach, however, is not linked to theories from learning science. In this paper, we focus on student-centered learning analytics (SCLA) i.e., the branch of HCLA that targets learners (Ochoa & Wise, 2020). We extend the HCID approach proposed in (Chatti et al., 2020a) by integrating SRL theories and models with the aim of providing a theoretically sound methodology to help LA researchers and developers systematically design and develop LA indicators for and with learners to effectively support their SRL activities. In the next sections, we provide the conceptual details of the proposed methodology and put it into practice through a concrete case study in a higher education context.

## 3.      THE SCLA-SRL METHODOLOGY

The Student-Centered Learning Analytics-enhanced Self-Regulated Learning (SCLA-SRL) methodology is a learner-first approach to indicator design, that starts with an understanding of the learners' real needs and goals and then designs indicators that best address these needs and goals. The methodology integrates principles, practices, and procedures required to carry out systematic research on LA-enhanced SRL. It is consistent with established theoretical frameworks in SRL, HCI, and InfoVis and provides a theory-driven process model for designing and developing LA indicators and dashboards that support SRL. The final objective of the SCLA-SRL methodology is to give answers to the following questions: Which LA indicators are needed to support SRL? and How to systematically design these indicators? This is achieved through two cyclical processes aiming at (1) **understanding** learners' needs and expectations from LA-enhanced SRL and (2) **empowering** learners to take control over the indicator design process, as depicted in Figure 1.

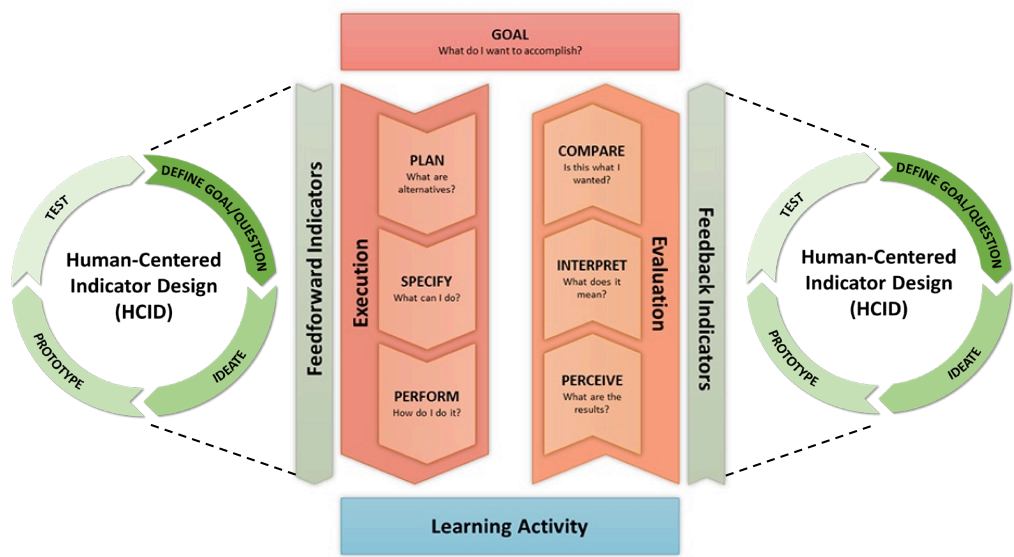

Figure 1. The SCLA-SRL Methodology (Understand & Empower)

## 3.1      Understand: Feedforward and feedback indicators for SRL

Which LA indicators are needed to support SRL? It is obvious that in an SRL scenario, the set of required indicators is unpredictable, because the indicators depend on the context, and different learners have different needs. It is therefore important to conduct qualitative user studies (e.g. interviews, focus groups) with learners to **understand** what they expect from LA-enhanced SRL and what are the LA indicators that they really need to support their SRL activities. Getting the right set of indicators would require asking the right set of questions. The different phases of the SRL process can provide a systematic way to ask the right set of questions and categorize the required LA indicators (Chatti & Muslim, 2019).

SRL is a cyclical process, composed of three general phases: (a) *goal setting*, (b) *executing*, and (c) *evaluating* (see Section 2.1). In an HCI context, Norman (2013) discusses seven stages of action that provide a guideline for developing usable and understandable new products or services, following a human-centered design (HCD) approach. By associating the typical three phase SRL model and Norman's seven stages of the action cycle, the SRL process can be modeled as a cyclical seven stages activity, as shown in the middle part of Figure 1. In detail, there are three major phases to an SRL activity: *goal setting*, *executing*, and *evaluating*. The execution phase is further subdivided into three stages that follow from the goal: *plan*, *specify*, and *perform*. The evaluation phase is further broken down into three stages: *perceive*, *interpret*, and *compare*.

The SRL activity cycle starts from the top with the learning goal (goal) and then goes through the three stages of execution: planning the possible learning activities to achieve those goals (plan), specify a learning activity path (specify), and perform the learning activity (perform). The cycle then goes through the three stages of evaluation: perceiving the results of the learning activity (perceive), trying to make sense of it (interpret), and comparing the learning outcome with the goal (compare). It is important to stress that most SRL activities require multiple feedback loops in which goals lead to subgoals, and the results of one activity are used to trigger further ones. Moreover, SRL activities do not always have to include all stages, nor do they have to proceed in a linear manner across all stages.

Each of the seven stages represents a possible question to ask towards an SRL activity. The seven-stage SRL activity cycle provides a useful tool for guiding the design of indicators for SRL. The role of LA is to help learners by conveying the information required to answer the learner's question at each stage of the execution and evaluation phases through appropriate indicators. Indicators that provide information that helps answer questions of execution (the left side of the middle part of Figure 1) are **feedforward indicators**. These include indicators for planning, awareness, and recommendation. Indicators providing information that aids in answering questions of evaluation (the right side of the middle part of Figure 1) are **feedback indicators**. These include indicators for self-monitoring, self-reflection, assessment, feedback, and motivation. The use of appropriate indicators at each stage enhances the overall SRL process. In the following, we summarize the questions related to the stages of the execution and evaluation phases along with the description of the indicators needed to answer these questions:

- *Goal* (What do I want to accomplish?): Provide information about the defined goals of the learning activity.
- *Plan* (What are alternatives?): Provide information needed to understand the possible actions that can be taken in order to reach the goals.
- *Specify* (What can I do?): Provide information to help learners decide on the appropriate learning activity path.
- *Perform* (How do I do it?): Provide information on best strategies in order to perform a task in an effective and efficient way.
- *Perceive* (What are the results?): Provide information to communicate the results of the performed tasks and the current state of the learning activity.

- *Interpret* (What does it mean?): Provide information to help learners understand the results and the impact of the learning activity in context.
- *Compare* (Is this what I wanted?): Provide information about progress towards goals.

The seven stages of the SRL activity cycle provide a guideline for developing structured interviews with learners to understand which indicators they really need to support their SRL activities. Rather than asking learners in an ad hoc manner about their abstract expectations of LA indicators, it is more effective to systematically ask about what they would do at each of the seven stages and then co-generate requirements for potential feedforward/feedback indicators that can support each stage.

## 3.2      Empower: Human-Centered Indicator Design

The question that might be raised now is: Once we have co-generated requirements for potential indicators, how to systematically co-design these indicators with learners? To get at this, we adopt the Human-Centered Indicator Design (HCID) approach proposed in (Chatti et al., 2020a). HCID brings together Norman's human-centered design (HCD) process (Norman, 2013) and Munzner's what-why-how visualization framework (Munzner, 2014) to provide a theory-informed approach for the systematic design of LA indicators. The main aim of HCID is to **empower** users to take control of the indicator design process in order to effectively meet their needs and goals. The HCID process encompasses four iterative stages: (1) Define Goal/Question, (2) Ideate, (3) Prototype, and (4) Test, as shown in the outer parts of Figure 1.

### 3.2.1      Define Goal/Question

The HCID process begins with a good understanding of users and the needs that the design is intended to meet. To achieve this, the initial stage in HCID is to define the Goal/Question to be addressed/answered by the indicator. These goals and questions are the results of the qualitative user study conducted based on the seven-stage SRL activity cycle, as discussed in the previous section.

### 3.2.2      Ideate

In the ideate stage, designers and learners come together to co-generate indicator ideas and concretize them in a systematic manner, using Indicator Specification Cards (ISC). An ISC describes a systematic workflow to get from the why? (i.e. user goal/question) to the how? (i.e. visualization). An example ISC is shown in Figure 3. It consists of two main parts, namely Goal/Question and Indicator. The Goal/Question part refers to the results of the previous stage of the HCID approach. The Indicator part is further broken down into three sub-parts, namely Task Abstraction (Why?), Data Abstraction (What?), and Idiom (How?), which reflect the three dimensions of Munzner's what-why-how visualization framework (Munzner, 2014).

### 3.2.3      Prototype

The next stage in the HCID process is to co-create indicator prototypes with learners based on the generated ISCs. The goal of this stage is to understand which of the visualization idioms proposed in the ideation stage are more effective for the tasks and data at hand. Paper or software prototypes can be used in this stage.

### 3.2.4    Test

The final stage in the HCID process is to get feedback from learners on the indicator prototypes. The aim is to verify that these prototypes effectively address/answer the goal/question defined in the first stage of the HCID process. Ideally, the evaluators should not be the same learners who participated in the previous HCID stages.

## 4.    CASE STUDY

To demonstrate the use of the SCLA-SRL methodology, we applied it to design LA indicators to support SRL activities of bachelor students attending an introductory Python programming course, offered in the winter semester 2019/20 at the University of Duisburg-Essen, Germany. This course was relevant for our study, as a high degree of self-regulation was required from students who were expected to plan and carry out their learning independently and also monitor and evaluate their progress throughout the course.

## 4.1    Understand

The first step was to follow the seven stages of the SRL activity cycle to understand learners' needs and goals and co-generate requirements for potential feedforward and feedback indicators that can support novice programmers' SRL. Authentic learning scenarios were constructed and validated to focus the interview conversation towards specific and realistic problems in the context of the programming course. An initial set of 14 scenarios were brainstormed by the authors together with tutors of the programming course, because they are familiar with the difficulties faced by the students in this course, such as lack of motivation and frustration, insufficient learning time, avoidance of help seeking, and lack of conceptual understanding. These scenarios covered the three phases of the SRL process: (a) goal setting (before learning), (b) executing (during learning), and (c) evaluating (after learning). The scenarios were constructed, tested, refined, and tested again in order to keep only the scenarios that are clear, desirable, feasible, measurable, and realistic from a student perspective. In order to observe the target group in their natural environment, we visited the exercise classes on different dates at the beginning of the semester and asked students (n=5) to give feedback on the scenarios in terms of the dimensions above by rating the scenarios on a five-point Likert scale from "Strongly disagree" to "Strongly agree". Each interview took between one and three hours, including introduction to the topic, presenting and discussing each of the 14 scenarios, and rating them. The resulting most important scenarios along with their refined final descriptions are summarized in Table 1.

   After the construction and validation of the learning scenarios, we conducted new interviews with other students from the same course (n=11) in order to co-generate requirements for potential feedforward and feedback indicators that can support these learning scenarios. The seven stages of the SRL activity cycle were used as a template to guide the interviews. All interviews were about 90 min long. The findings of this round of interviews has led to a first set of feedforward (FF) and feedback (FB) indicators that can be found in Table 1.

Table 1. SRL Scenarios and Potential Feedforward / Feedback Indicators

| SRL Phases | SRL Scenarios | Feedforward (FF) and Feedback (FB) Indicators |
|---|---|---|
| Before learning (G = Goal setting) | **G1**: Get familiar with the Python programming concepts discussed in the course<br><br>**G2**: Successfully solve the tasks in all assignment sheets to prepare for the exam | |
| During learning (Ex = Executing) | **Ex1**: You read the task on the assignment sheet. You believe that you have understood the task correctly. But you do not know how to solve it. You give yourself 15 minutes to try to understand it better | **FF1**, *Help Seeking*: As a feedforward indicator of whether help is needed to solve a task<br>**FF2**, *Task Strategies*: As a feedforward indicator for using promising strategies and avoiding others to solve a task<br>**FF3**, *Information Seeking*: As a feedforward indicator of which resources are suitable as references to solve a task<br>**FF4**, *Information Seeking*: As a feedforward indicator to reuse certain code<br>**FF5**, *Planning*: As a feedforward indicator of whether one should learn more about a topic or concept |
| | **Ex2**: While you try to solve a programming task, error messages appear. You want to fix the error messages to be able to solve the task. You give yourself 15 minutes to deal with the error | **FF6**, *Information Seeking*: As a feedforward indicator of which websites can be used to search for a solution<br>**FF7**, *Help Seeking*: As a feedforward indicator of how helpful was a specific website in solving a problem<br>**FF8**, *Help Seeking*: As a feedback indicator of how helpful was an exercise class in solving a problem<br>**FF9**, *Information Seeking*: As a feedforward indicator of which questions are suitable for finding answers to a problem |
| After learning (Ev = Evaluating) | **Ev1**: You have solved several programming tasks so far. Now you take 30 minutes to figure out which programming concepts (looping, recursion, sorting etc.) you still have problems with | **FB1**, *Learning Success*: As a feedback indicator of how adequately one has dealt with a problem<br>**FB2**, *Learning Difficulties*: As a feedback indicator of how difficult a task was perceived<br>**FB3**, *Learning Difficulties*: As a feedback indicator of which mistakes you make more often in which context<br>**FB4**, *Learning Success*: As feedback indicator for the distance to the learning goal G1<br>**FB5**, *Motivation*: As a feedback indicator of how strong is your motivation to learn a new concept<br>**FB6**, *Learning Success*: As a feedback indicator of how well one has understood a particular concept<br>**FB7**, *Learning Difficulties*: As a feedback indicator for which concepts one still has difficulties with |
| | **Ev2**: You have solved all assignment sheets so far and want to prepare for the exam. Now you take 30 minutes to figure out whether you have learned enough for the exam or whether you should continue learning | **FB8**, *Learning Success*: As feedback indicator for the distance to the learning goal G2 |

## 4.2     Empower

The next step was to use the requirements for the potential feedforward and feedback indicators as input to the HCID loop to systematically co-design these indicators. We organized two co-design workshops with the same group of students that were interviewed before. However, only six students (n = 6) were able to attend the workshops. The workshops were held online via an online video conferencing tool and were four hours long each.

The aim of the first workshop was to brainstorm together with students ideas for possible visualizations to illustrate the potential feedforward and feedback indicators and to co-develop paper prototypes for the indicators, based on Indicator Specification Cards (ISCs) (see Section 3.2.2). Figure 3 shows an example ISC to describe the feedforward indicator FF5. In practice, this step was too complex and at times confusing to the participants, since in general, bachelor students do not have a strong background in InfoVis theory. During this step we noticed how important the role of the moderator was. The moderator can provide explanation when participants are not aware of why specific visualizations are more effective than others or when someone needs more background information about the used what-why-how visualization framework. In addition, we summarized InfoVis design guidelines to advance students' understanding of the topic and help them choose the right visualization for the task and data at hand, as depicted in Figure 2. The participants found these illustrations helpful to learn more about InfoVis theory. Despite the complexity of this step, the ISCs enabled to get the indicators right by following state-of-the-art InfoVis design practices. Based on the ISCs, paper prototypes were produced with students (see Figure 3 for an example).

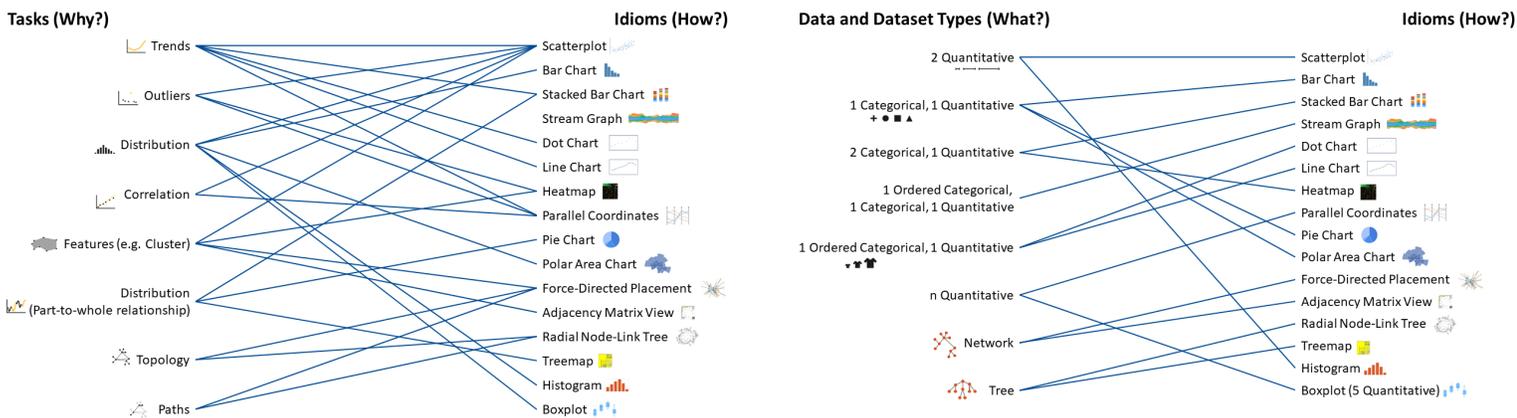

Figure 2. InfoVis Design Guidelines

The goal of the second workshop was to test these prototypes regarding the effectiveness of the visualizations to illustrate the related indicators. To prepare for this workshop, the paper prototypes were converted to Figma software prototypes by the authors (see Figure 3). To evaluate the initial prototypes, the students were asked for their feedback first on the importance of the different indicators to meet the learning goals G1 and G2 and second on the usefulness of the associated prototypes and how to improve them. Feedback addressed for example the need to make simple visualizations (e.g. bar charts, tables) and to remove some of the prototypes and combine others. This resulted in nine prototypes (FF1, FF5, FB1, FB2, FB3, FB4, FB6, FB7, FB8) that were perceived to be effective for G1 and G2 and were thus adopted for the next round. Feedback has been used to improve the design of the indicators and merge some of them, as well as to adapt the arrangement of the indicators in a dashboard by following an overview-detail approach. Figure 4 and Figure 5 show the final dashboard with the various indicators that have been developed. The first two indicators use bar chart and concept map to provide an overview of the learning progress at an abstract level, allowing students to monitor their progress (FB4), see the relationships between the concepts, and plan their next learning activities (FF5). The next three indicators use table and heatmap to provide more details at the concept, task, and solution levels to help students monitor their progress towards learning goals at a more granular level and assess their learning success (FB1, FB6) and difficulties (FB2, FB3, FB7), and eventually seek help (FF1). The last two indicators use bar chart and table to help students monitor their learning performance to prepare for the exam (FB8, FF5).

| Goal/Question | |
|---|---|
| **Goal:** Plan the next learning activity | |
| **Question:** What are the programming concepts that I need to learn for this course? How do they relate to each other? What is the next concept that I need to learn about? | |
| **Indicator** | |
| **Task Abstraction (Why?)** *Identify* the concept structure (*topology*) of the course *Identify* relationships (*features*) between the course concepts *Identify* possible learning paths | |
| **Data Abstraction (What?)** | |
| ***Raw data*** | ***Derived data*** |
| Course data: • Concepts (categorical attribute) • Concept mastering level (categorical attribute) | Course tree: • Nodes: Concepts • Links: Sub-concept-of |
| **Idiom (How?)** | |
| ***How to encode?*** | ***How to interact?*** |
| Radial Node-Link Tree Color | *Manipulate:* select (concept), zoom, pan |

**(a)**

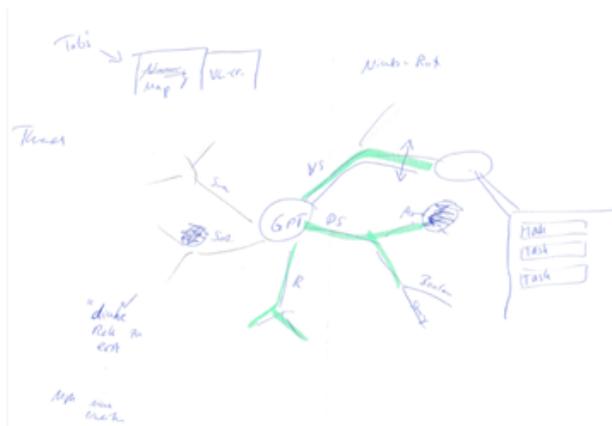 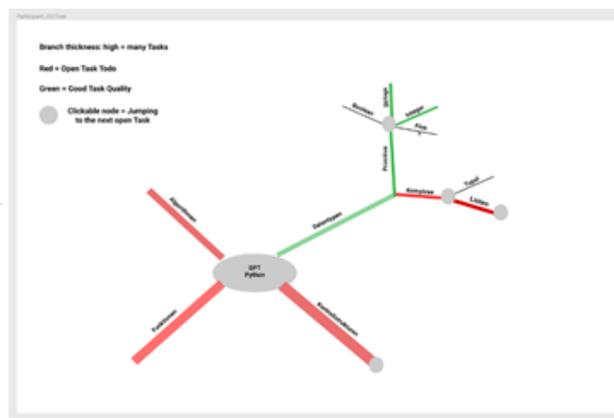

**(b)**                                        **(c)**

Figure 3. (a) Indicator Specification Card (ISC), (b) Paper Prototype, and (c) Software Prototype for Feedforward Indicator FF5

## Overview

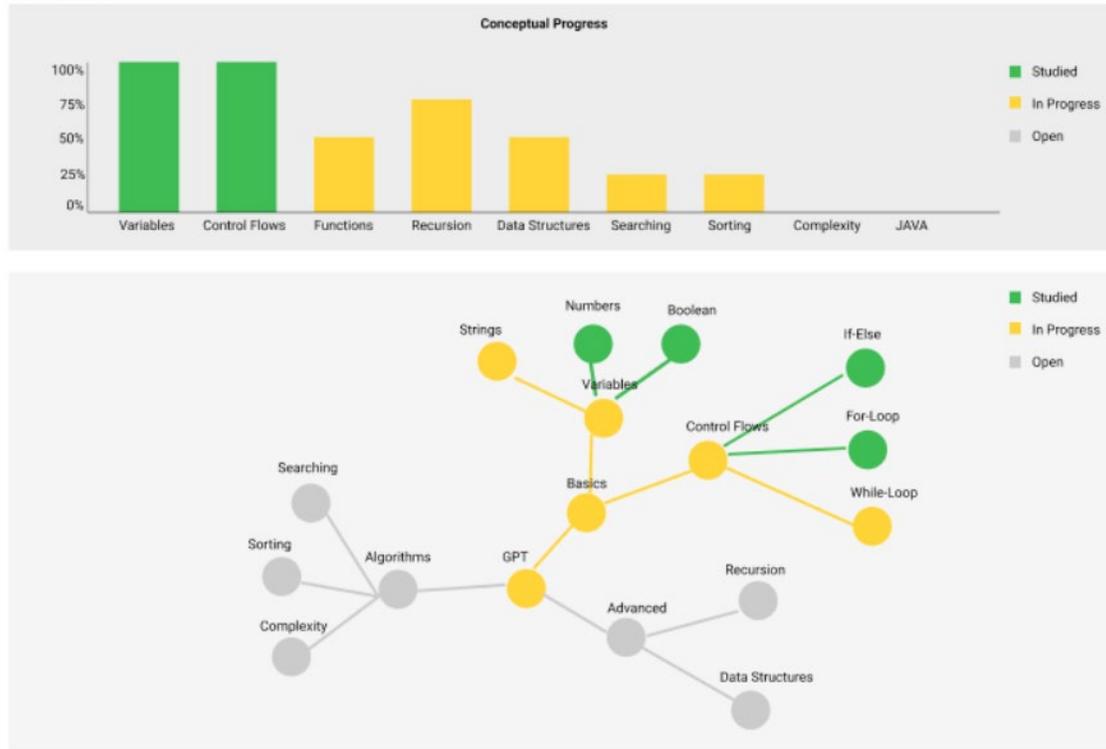

## Concepts

| ID | Name | Tasks Solved | Current Lvl of Understanding | Status | Priority | | |
|----|------|--------------|------------------------------|--------|----------|---|---|
| 1 | Variables | 15 | ★ ★ ★ ★ ★ | 🟢 Studied | 1 | ✏ | 🗑 |
| 2 | Arrays | 20 | ★ ★ ★ | 🟡 In Progress | 3 | ✏ | 🗑 |
| 3 | Is-Else-Statements | 14 | ★ ★ ★ ★ | 🟢 Studied | 1 | ✏ | 🗑 |
| 4 | For-Loops | 6 | ★ ★ ★ | 🟡 In Progress | 3 | ✏ | 🗑 |
| 5 | While-Loops | 7 | ★ ★ | 🟡 In Progress | 4 | ✏ | 🗑 |
| 6 | Functions | 2 | ★ ★ | 🟡 In Progress | 4 | ✏ | 🗑 |
| 7 | Recursion | 1 | ★ | ⚪ Open | 5 | ✏ | 🗑 |

Figure 4. Final Prototypes (Part 1)

Figure 5. Final Prototypes (Part 2)

# 5.      CONCLUSION

Learning analytics (LA) is opening up new opportunities for promoting self-regulated learning (SRL). In this paper, we stressed the need for a methodology to serve as a commonly accepted framework for LA-enhanced SRL research and a template for the systematic design of LA indicators to support SRL. We argued that for LA dashboards and indicators to be accepted and adopted by learners, their design and development need to be much more linked to sound theories and models in learning science, human-computer interaction (HCI), and information visualization (InfoVis). Moreover, learners have to be involved throughout the whole design process. To get at this, we proposed the Student-Centered Learning Analytics-enhanced Self-Regulated Learning (SCLA-SRL) methodology to guide the systematic design of LA indicators that support SRL. We then presented a case of applying this methodology to design and develop LA indicators to support novice programmers' SRL in a higher education context. We expect that this case study will provide a useful template for LA researchers and developers who want to apply SCLA-SRL to their efforts. The novelty of the SCLA-SRL methodology resides in the fact that it instantiates design guidelines from the HCI and InfoVis areas and combines them with SRL models. This effort contributes to LA research by providing a theoretically sound framework for successfully carrying out LA-enhanced SRL research and a template to systematically design LA indicators and dashboards that support SRL.